\documentclass[10pt,conference]{IEEEtran}
\IEEEoverridecommandlockouts
\usepackage{geometry}
\usepackage{graphicx}
\usepackage{epstopdf}
\usepackage{cite}
\usepackage{amsmath,amssymb,amsfonts}
\usepackage{algorithmic}
\usepackage{float}
\usepackage{graphicx}
\usepackage{textcomp}
\usepackage{xcolor}
\usepackage{array}
\usepackage{multicol}
\usepackage{multirow}
\usepackage{makecell}
\allowdisplaybreaks
\usepackage{subcaption}
\usepackage[ruled,vlined]{algorithm2e}
\usepackage{bm}
\usepackage{booktabs}
\usepackage{array}
\usepackage{cancel}
\usepackage[normalem]{ulem}

\newtheorem{proposition}{Proposition}

\def\BibTeX{{\rm B\kern-.05em{\sc i\kern-.025em b}\kern-.08em
    T\kern-.1667em\lower.7ex\hbox{E}\kern-.125emX}}
\begin{document}

\title{Distributed Nonlinear Transform Source-Channel Coding for Wireless Correlated Image Transmission
\thanks{The code of this work will be released at https://github.com/SJTUmxtao/D-NTSCC after publication.}
\thanks{This work was supported by National Science and Technology Major Project - Mobile Information Networks under grant 2024ZD1300700, the NSF of China under grant 62125108, and the Science and Technology Commission Foundation of Shanghai under grant  24DP1500702.}
}

\makeatletter
\newcommand{\linebreakand}{%
  \end{@IEEEauthorhalign}
  \hfill\mbox{}\par
  \mbox{}\hfill\begin{@IEEEauthorhalign}
}
\makeatother
\author{Yufei Bo and Meixia Tao\\
Department of Electronic Engineering and the Cooperative Medianet Innovation Center (CMIC) \\Shanghai Jiao Tong University, China\\
Emails: \{boyufei01, mxtao\}@sjtu.edu.cn

}
\maketitle

\begin{abstract}

This paper investigates distributed  joint source-channel coding (JSCC) for correlated image semantic transmission over wireless channels. 
In this setup, correlated images at different transmitters are separately encoded and transmitted through dedicated channels for joint recovery at the receiver. 
We propose a novel distributed nonlinear transform source-channel coding (D-NTSCC) framework. 
Unlike existing learning-based approaches that implicitly learn source correlation in a purely data-driven manner, our method explicitly models the source correlation through joint distribution.
Specifically, the correlated images are separately encoded into latent representations via an encoding transform function, followed by a JSCC encoder to produce channel input symbols. 
A learned joint entropy model is introduced to determine the transmission rates, which more accurately approximates the joint distribution of the latent representations and captures source dependencies, thereby improving rate-distortion performance. 
At the receiver, a JSCC decoder and a decoding transform function reconstruct the images from the received signals, each serving as side information for recovering the other image. 
Therein, a transformation module is designed to align the latent representations for maximal correlation learning. 
Furthermore, a loss function is derived to jointly optimize encoding, decoding, and the joint entropy model, ensuring that the learned joint entropy model approximates the true joint distribution. 
Experiments on multi-view datasets show that D-NTSCC outperforms state-of-the-art distributed schemes, demonstrating its effectiveness in exploiting source correlation.

\end{abstract}

\section{Introduction}

Semantic communications, or \textit{task-oriented communications}, emerge as a promising paradigm for future 6G networks, focusing on transmitting only the essential information directly relevant to the tasks at the receiver\cite{gunduz2022beyond}. 
Leveraging deep learning, semantic communications replace traditional hand-crafted communication modules by neural network (NN)-based counterparts. 
Such systems typically employ a neural semantic encoder for joint source-channel coding (JSCC), directly mapping source data to channel symbols for transmission.
The receiver employs a neural semantic decoder to execute desired tasks.
The whole system is optimized end-to-end, taking into account source distribution, channel distribution, as well as receiver tasks\cite{dai2022nonlinear, 10495330}.  
Notably, compared with traditional Shannon-type communications, semantic communications offer significant gains in transmission efficiency and downstream task performance.
This work investigates multi-user semantic communications where distributed transmitters with correlated image sources communicate with a common receiver. 
The task at the receiver is to reconstruct the images with the highest fidelity at minimum channel usage. 
A typical application scenario is multi-view surveillance systems, where multiple separately located cameras with probably overlapping fields of view capture images and transmit them to a center node. 
Traditionally, this problem is formulated as distributed source coding (DSC).
As demonstrated by the Slepian-Wolf theorem\cite{slepian1973noiseless}, separate encoding and joint decoding of correlated sources can achieve the same compression rate as joint encoding-decoding under lossless compression. 
This theorem has then been extended to the lossy case by Berger\cite{berger1978multiterminal} and Tung\cite{tung1978multiterminal}.
Building upon these information theoretical results, many practical DSC schemes have been proposed, including conventional hand-crafted approaches such as \cite{pradhan2003distributed, yang2008multiterminal} and data-driven ones such as \cite{diao2020drasic, ldmic}.
Note that these distributed coding methods are limited to source coding and do not account for the characteristics of the communication channel.

\begin{figure*}
    \centering
    \includegraphics[scale=0.52]{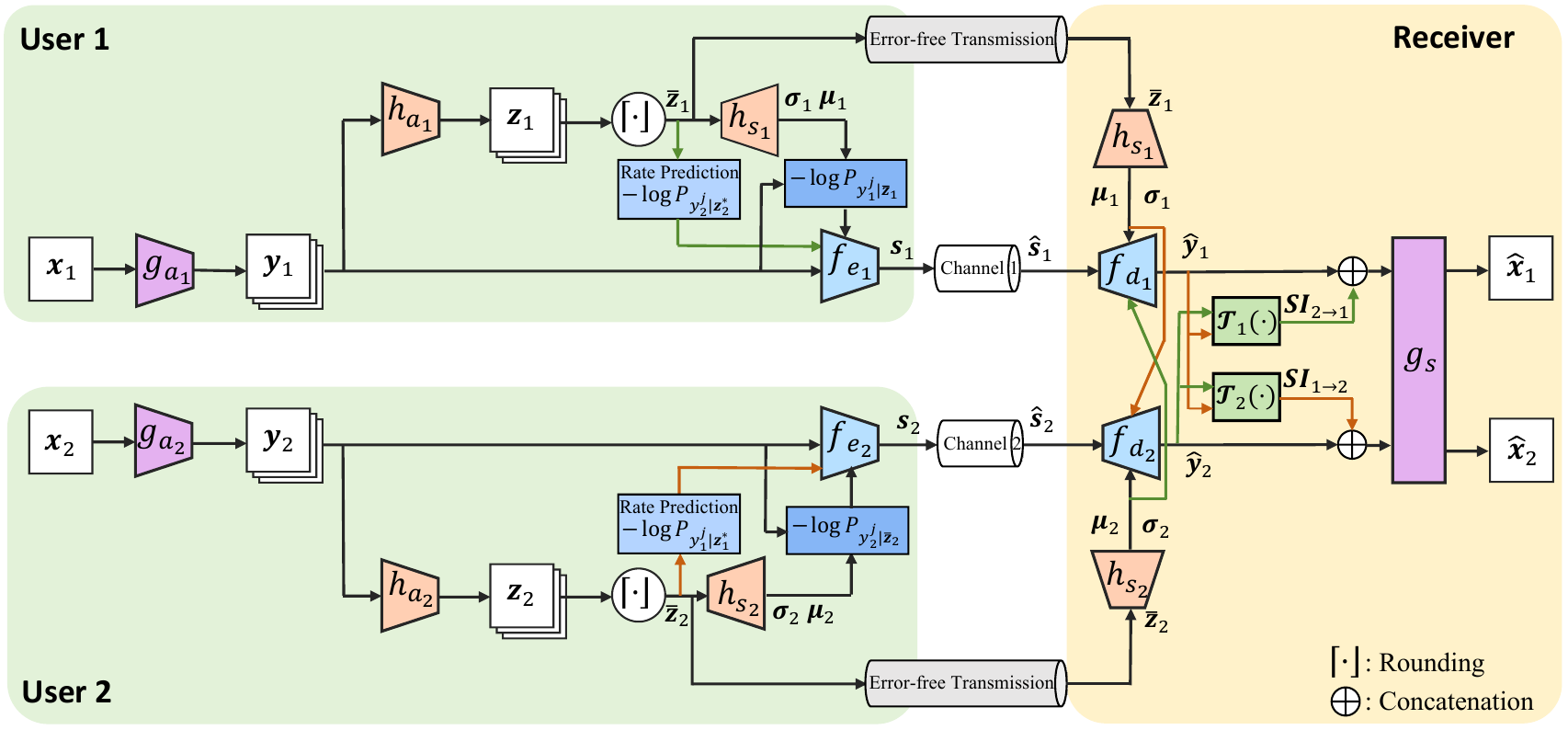}
    \caption{Overall architecture of the proposed D-NTSCC scheme.}
    \label{system model}
    \vspace{-0.6cm}
\end{figure*}

Recently, several deep learning-based distributed JSCC methods have emerged \cite{wang2022distributed, li2024content}.
The work \cite{wang2022distributed} employs an autoencoder-based architecture with two lightweight edge encoders and a central decoder, introducing a signal-to-noise-ratio (SNR)-aware cross-attention (CA) module at the receiver for enhanced feature interaction.
Additionally, the work \cite{li2024content} introduces a multi-layer CA mechanism to fuse image features across different pixel levels for better correlation exploitation.
However, existing approaches rely on implicit learning of the source correlation in a data-driven manner, prompting the need for further research into more effective strategies to exploit source correlation.

In this paper, we propose distributed nonlinear transform source-channel coding (D-NTSCC), a novel framework that exploits source correlation via joint distribution while retaining advantages of nonlinear transform coding for JSCC\cite{dai2022nonlinear}. 
Specifically, multiple correlated image sources are separately encoded: first into latent representations by encoding transform functions, and then into channel input symbols through JSCC encoders.
These symbols are transmitted through independent noisy channels to a common receiver. 
The transmission rate of the channel input symbols is determined by the code rate of the latent representation, which is approximated by a learned joint entropy model.
At the receiver, a JSCC decoder recovers the latent representation, and a joint decoding transform function restores the images, with each image utilizing other noisy latent representations as side information.

Three key innovations are made to better exploit source correlation.
First, a joint entropy model is developed to approximate the joint distribution of the latent representations, capturing their dependencies more accurately and improving rate-distortion performance.
Second, the loss function for D-NTSCC is derived using variational inference, which jointly learns encoding, decoding and the joint entropy model, thereby ensuring the learned entropy model approximates the true joint distribution.
Third, we propose a transformation module at the receiver to locate and transform the most relevant side information in the other latent representation to an expected pose, ensuring maximal correlation utilization in subsequent layers.

Experiments based on real-world multi-view datasets validate the advantages of the D-NTSCC scheme.
Compared to existing distributed deep JSCC coding schemes and the point-to-point nonlinear transform source-channel coding (NTSCC) baseline scheme, the proposed method is superior across all evaluated transmission rates, achieving the highest performance on both pixel-wise and perceptual metrics.

\section{System Model and Overall Architecture}

\begin{figure*}
    \begin{align}
        &\ \ \ \ \ \mathbb{E}_{p_{\mathbf{x}_1,\mathbf{x}_2}}D_{KL}[q(\mathbf{\hat{s}}_1,\mathbf{\hat{s}}_2,\mathbf{\tilde{z}}_1,\mathbf{\tilde{z}}_2|\mathbf{x}_1,\mathbf{x}_2)||p(\mathbf{\hat{s}}_1,\mathbf{\hat{s}}_2,\mathbf{\tilde{z}}_1,\mathbf{\tilde{z}}_2|\mathbf{x}_1,\mathbf{x}_2)]\nonumber\\
        &=\mathbb{E}_{p_{\mathbf{x}_1,\mathbf{x}_2}}\mathbb{E}_{q_{\hat{\mathbf{s}}_1,\hat{\mathbf{s}}_2,\tilde{\mathbf{z}}_1,\tilde{\mathbf{z}}_2|\mathbf{x}_1,\mathbf{x}_2}} \Bigl\{ 
        \underbrace{-\log p(\mathbf{x}_1,\mathbf{x}_2|\mathbf{\hat{s}}_1,\mathbf{\hat{s}}_2,\mathbf{\tilde{z}}_1,\mathbf{\tilde{z}}_2)}_{\text{weighted distortion}} 
        -\underbrace{\log p(\mathbf{\hat{s}}_1|\mathbf{\tilde{z}}_1)-\log p(\mathbf{\hat{s}}_2|\mathbf{\tilde{z}}_2)-\log p(\mathbf{\tilde{z}}_1,\mathbf{\tilde{z}}_2)}_{\text{rate}} 
       \Bigr\} + \text{const}
        \label{kl}
    \end{align}
    \hrulefill
    \vspace{-0.5cm}
\end{figure*}


We consider semantic communications in a distributed setting,
where two physically separated terminals send correlated but not necessarily independently distributed image sources to a common receiver through independent channels.  
Here two source terminals are considered for simplicity. 
The extension to more than two sources can be done easily. 
Let the two images be denoted as $\mathbf{x}_1, \mathbf{x}_2\in \mathbb{R}^{C\times H\times W}$, where $C$, $H$, and $W$ represent the channel, height, and width of the images, respectively.

The overall architecture of the proposed D-NTSCC is shown in Fig.~\ref{system model}. 
Specifically, each transmitter $i$, for $i\in \{1,2\}$, first employs an NN-based analysis transform function $g_{a_i}$ with parameters $\boldsymbol{\phi}_{g_i}$ to map $\mathbf{x}_i$ into a latent representation $\mathbf{y}_i\in\mathbb{R}^{N_y}$.
Then, a JSCC encoder $f_{e_i}$ with parameters $\boldsymbol{\phi}_{f_i}$ encodes $\mathbf{y}_i$ into channel input $\mathbf{s}_i$, which is reshaped into a complex-valued vector $\mathbf{s}_i\in\mathbb{C}^{n}$ for transmission.
Here, $n$ denotes the number of channel use. Each element of $\mathbf{s}_i$ is subject to the average transmit power constraint 
$P$, which is assumed to be the same for both transmitters.

Furthermore, to achieve adaptive rate transmission, a hyperprior variable $\mathbf{z}_i$ is introduced to characterize the probability density function (PDF) of $\mathbf{y}_i$, thereby determining the channel bandwidth cost for its each dimension.
A parametric analysis transform $h_{a_i}$ maps $\mathbf{y}_i$ to the hyperprior $\mathbf{z}_i\in\mathbb{R}^{N_z}$, where $N_z\ll N_y$.
The hyperprior $\mathbf{z}_i$ represents the dependencies among the elements of $\mathbf{y}_i$, allowing them to be assumed independent given $\mathbf{z}_i$\cite{balle2018variational}.
Additionally, it summarizes the means and standard deviations of $\mathbf{y}_i$, and is quantized, compressed, and transmitted to the receiver.
Specifically, each element $y_i^j$ of $\mathbf{y}_i$, for $j=1,...,N_y$, is modeled as a Gaussian variable with mean $\mu_i^j$ and standard deviation $\sigma_i^j$, which are estimated by applying a parametric synthesis transform $h_{s_i}$ to the quantized $\mathbf{\bar{z}}_i$.
Note that, to ensure training stability and improve density matching, both $\mathbf{y}_i$ and $\mathbf{z}_i$ are convolved with a standard uniform density when modeling their probability distributions, denoted as $\mathbf{\tilde{y}}_i$ and $\mathbf{\tilde{z}}_i$, respectively.
Therefore, the probability of $\mathbf{\tilde{y}}_i$ given $\mathbf{\tilde{z}}_i$ can be written as
\begin{equation}
p(\mathbf{\tilde{y}}_i|\mathbf{\tilde{z}}_i)=\prod_j \left ( \mathcal{N}(\mu_i^j, \sigma_i^j)*\mathcal{U}(-\frac{1}{2}, \frac{1}{2}) \right )(\tilde{y}_i^j),
\end{equation}
with
\begin{equation}
    \boldsymbol{\sigma}_i, \boldsymbol{\mu}_i =h_{s_i}(\mathbf{\tilde{z}}_i;\boldsymbol{\theta}_{h_{i}}),
\end{equation}
where $*$ denotes the convolution operation, and $\boldsymbol{\theta}_{h_i}$ encapsulates the parameters of $h_{s_i}$.
Therefore, the code rate required to transmit $\mathbf{y}_i$ is given by $-\log p(\mathbf{\tilde{y}}_i|\mathbf{\tilde{z}}_i)$, which is measured in bits when the logarithm is base-2.
The transmission rate of the channel input $\mathbf{s}_i$ is constrained proportionally to this value.
Moreover, since the hyperpriors contain transmission rate information, they are also required at the receiver for decoding. Assuming error-free transmission, we estimate their transmission cost using their joint density model $p(\mathbf{\tilde{z}}_1, \mathbf{\tilde{z}}_2|\boldsymbol{\psi})$, where $\boldsymbol{\psi}$ denotes the distribution parameters.
Details of this joint density model are provided in Section III-B.

A dedicated and independent AWGN channel for each transmitter is considered.
The received vector can be written as
$ \mathbf{\hat s}_i = \mathbf{s}_i + \mathbf{n}_i $, 
where $\mathbf{n}_i\sim \mathcal{CN}(\mathbf{0}, \epsilon^2_i\mathbf{I}_{n\times n})$ is the i.i.d. complex Gaussian noise with zero mean and variance $\epsilon^2_i$.
For simplicity, the two noise variances are assumed to be same, \emph{i.e.,} $\epsilon^2_1=\epsilon^2_2=\epsilon^2$.
Thus, the two channels have the same SNR, which is defined as $\frac{P}{\epsilon^2}$.


At the receiver, the received vector $\mathbf{\hat s}_i$ ($i\in\{1,2\}$) first passes through a JSCC decoder $f_{d_i}$ with parameters $\boldsymbol{\theta}_{f_i}$ to recover the latent representation, denoted as $\mathbf{\hat{y}}_i$, with the assistance of $\mathbf{\bar{z}}_1$ and $\mathbf{\bar{z}}_2$.
Then, a synthesis transform function $g_s$ with parameters $\boldsymbol{\theta}_{g}$ performs joint decoding, recovering $\mathbf{\hat{x}}_1$ and $\mathbf{\hat{x}}_2$ from $\mathbf{\hat{y}}_1$ and $\mathbf{\hat{y}}_2$.
During this process, the two recovered latent representations serve as side information for each other, which can be written as
\begin{align}
    \mathbf{\hat x}_1 &= g_s(\mathbf{\hat y}_1|\mathcal{T}_1(\mathbf{\hat y}_2)),\\
    \mathbf{\hat x}_2 &= g_s(\mathbf{\hat y}_2|\mathcal{T}_2(\mathbf{\hat y}_1)),
\end{align}
where $\mathcal{T}_1(\cdot)$ is a transformation module that spatially aligns $\mathbf{\hat y}_2$ with $\mathbf{\hat y}_1$, enabling the subsequent NN layers to better utilize their correlation. 
The transformed $\mathbf{\hat y}_2$, namely, $\mathcal{T}_1(\mathbf{\hat y}_2)$, is denoted as $\mathbf{SI}_{2\rightarrow 1}$.
The same applies to $\mathcal{T}_2(\cdot)$.
Details of this module shall be introduced in Section IV-C.

\section{Optimization with Variational Inference}

\subsection{Loss Function Derivation}

The optimization problem of nonlinear transform coding is typically formulated as a variational autoencoder (VAE), where the synthesis transform corresponds to the generative model of the VAE while the analysis transform is linked to the inference model\cite{balle2018variational}.
In our setting, the goal of variational inference is to approximate the true but intractable joint posterior $p(\mathbf{\hat{s}}_1,\mathbf{\hat{s}}_2,\mathbf{\tilde{z}}_1,\mathbf{\tilde{z}}_2|\mathbf{x}_1,\mathbf{x}_2)$ with a parametric variational density $q(\mathbf{\hat{s}}_1,\mathbf{\hat{s}}_2,\mathbf{\tilde{z}}_1,\mathbf{\tilde{z}}_2|\mathbf{x}_1,\mathbf{x}_2)$ by minimizing their Kullback-Leibler (KL) divergence, which is given by the following proposition.

\begin{proposition}
\label{prop}
The expectation of the KL divergence between $q(\mathbf{\hat{s}}_1,\mathbf{\hat{s}}_2,\mathbf{\tilde{z}}_1,\mathbf{\tilde{z}}_2|\mathbf{x}_1,\mathbf{x}_2)$ and  $p(\mathbf{\hat{s}}_1,\mathbf{\hat{s}}_2,\mathbf{\tilde{z}}_1,\mathbf{\tilde{z}}_2|\mathbf{x}_1,\mathbf{x}_2)$ over the data distribution $p_{\mathbf{x}_1,\mathbf{x}_2}$ is given by \eqref{kl}, where \textit{const} denotes a constant.
\end{proposition}

The proof can be found in Appendix A.
Examining each term in \eqref{kl} more closely, we find that the first term corresponds to the distortion of the correlated sources. 
Minimizing this term is equivalent to minimizing the expected distortion of the reconstructed images.
Moreover, the probability  $p(\mathbf{\hat{s}}_i|\mathbf{\tilde{z}}_i)$ is the convolution of $p(\mathbf{s}_i|\mathbf{\tilde{z}}_i)$ and the Gaussian noise.
Therefore, the second and third terms can be interpreted as the cost of transmitting $\mathbf{s}_1$ and $\mathbf{s}_2$, respectively, which are proportional to the rates $-\log p(\mathbf{\tilde{y}}_1|\mathbf{\tilde{z}}_1)$ and $-\log p(\mathbf{\tilde{y}}_2|\mathbf{\tilde{z}}_2)$.
The fourth term corresponds to the code rate of the hyperpriors, reflecting the cost of transmitting $(\mathbf{\tilde{z}}_1, \mathbf{\tilde{z}}_2)$.

Based on Proposition 1 and the discussion above, we thus define the loss function of the D-NTSCC scheme as
\begin{align}
\mathcal{L} &= \mathbb{E}_{p_{\mathbf{x}_1, \mathbf{x}_2}} \Big[ 
    d(\mathbf{x}_1, \mathbf{\hat{x}}_1) + d(\mathbf{x}_2, \mathbf{\hat{x}}_2) \nonumber \\
    + \lambda \big( -\eta 
    &\log p(\mathbf{\tilde{y}}_1|\mathbf{\tilde{z}}_1) 
    - \eta \log p(\mathbf{\tilde{y}}_2|\mathbf{\tilde{z}}_2) -\log p(\tilde{\mathbf{z}}_1,\mathbf{\tilde{z}}_2)\big) 
\Big],
\label{loss}
\end{align}
where $d(\cdot,\cdot)$ represents the distortion measure between the source $\mathbf{x}_i$ and its recovery $\mathbf{\hat{x}}_i$, $\lambda$ controls the trade-off between rate and distortion, and $\eta$ indicates the proportional relationship between the code rate of the latent representation and the transmission rate of the channel input.

\subsection{Joint Probability Function of Hyperprior Variables}

In this subsection, we present the code rate estimation of the hyperpriors $(\mathbf{z}_1, \mathbf{z}_2)$ by defining their joint probability function.  
Note that for univariate probability modeling, Ball\'{e} \textit{et al.}\cite{balle2018variational} used a method based on the cumulative distribution function (CDF).
By ensuring that an NN-parameterized function $F(x):\mathbb{R}\rightarrow [0,1]$ satisfies the properties of a valid CDF — Specifically, $F(-\infty)=0$, $F(+\infty)=1$, and $\frac{\partial F(x)}{\partial x}\ge 0$ — this function can be utilized to model the probability distribution of the univariate hyperprior.
However, applying this method to bivariate variables is challenging.
Particularly, it is difficult to ensure that the second-order mixed partial derivative of an NN-parameterized function remains non-negative, which is necessary for maintaining the properties of a valid joint CDF.

\begin{figure}[t]
    \centering
    \begin{subfigure}[b]{0.45\textwidth}
        \centering
        \includegraphics[width=\textwidth]{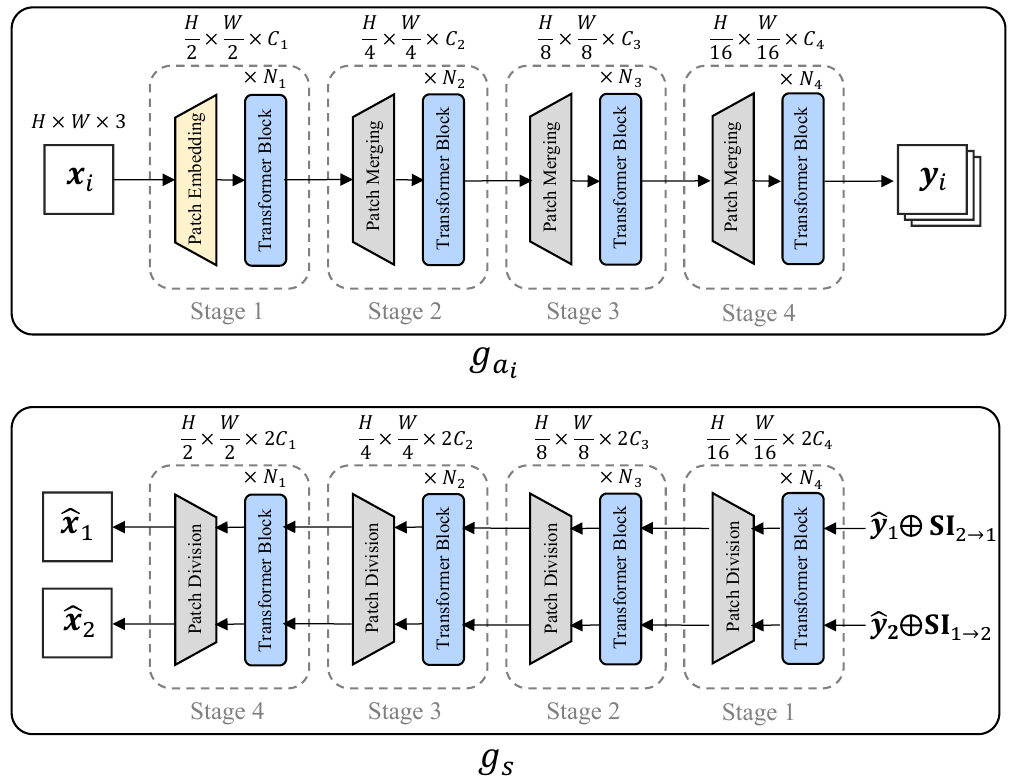}
        \caption{Nonlinear transforms $g_{a_i}$ and $g_s$.}
    \end{subfigure}
    \hfill
    \begin{subfigure}[b]{0.45\textwidth}
        \centering
        \includegraphics[width=\textwidth]{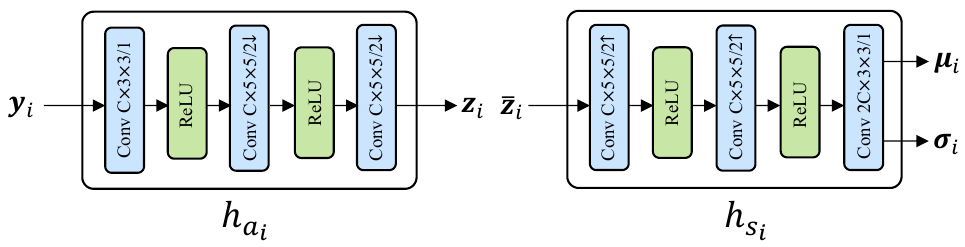}
        \caption{Analysis transform $h_{a_i}$ and synthesis transform $h_{s_i}$.}
    \end{subfigure}
    \caption{NN architectures of nonlinear transform modules.}
    \label{ntc modules}
    \vspace{-0.6cm}
\end{figure}

Therefore, we propose a bivariate probability modeling method based on PDF.
We model $(\mathbf{z}_1,\mathbf{z}_2)$ using a parametric, pairwise factorized density model, where each pairwise joint density $p(z_1^j,z_2^j)$ is parameterized as a Gaussian Mixture Model (GMM). 
Specifically, the density is expressed as
\begin{equation}
\setlength\abovedisplayskip{4pt}
\setlength\belowdisplayskip{2pt}
    p(\mathbf{z}_1,\mathbf{z}_2)=\prod_j p(z_1^j,z_2^j),
\end{equation}
where $p(z_1^j,z_2^j)$ is defined as a mixture of Gaussians
\begin{equation}
\setlength\abovedisplayskip{4pt}
\setlength\belowdisplayskip{5pt}
    p(z_1^j,z_2^j)=\sum_{k=1}^K \pi_k^j \mathcal{N}(\mathbf{m}_k^j,\boldsymbol{\Sigma}_k^j)(z_1^j,z_2^j),
\end{equation}
with $\pi_k^j$, $\mathbf{m}_k^j$, and $\boldsymbol{\Sigma}_k^j$ representing the mixture weights, means, and covariances of the $k$-th Gaussian component of the $j$-th element pair $(z_1^j,z_2^j)$, respectively,  and $K$ representing the number of mixtures.
The parameters $\pi_k^j$, $\mathbf{m}_k^j$, and $\boldsymbol{\Sigma}_k^j$, encapsulated by $\boldsymbol{\psi}$, are learnable parameters optimized by NNs.
Furthermore, the density $p(\mathbf{\tilde{z}}_1,\mathbf{\tilde{z}}_2)$ is obtained by convolving $p(\mathbf{z}_1,\mathbf{z}_2)$ with a standard uniform density to approximate the density of the quantized hyperpriors $(\mathbf{\bar{z}}_1,\mathbf{\bar{z}}_2)$, and can be written as
\begin{equation}
    \setlength\abovedisplayskip{4pt}
\setlength\belowdisplayskip{5pt}
p(\mathbf{\tilde{z}}_1,\mathbf{\tilde{z}}_2)=\prod_j\left [\sum_{k=1}^K \pi_k^j \left(\mathcal{N}(\mathbf{m}_k^j,\boldsymbol{\Sigma}_k^j)*\mathcal{U}(-\frac{1}{2}, \frac{1}{2})\right)(\tilde{z}_1^j,\tilde{z}_2^j)\right].
\end{equation} 
Finally, the probability at integer points is obtained by integrating the density $p(\mathbf{\tilde{z}}_1,\mathbf{\tilde{z}}_2)$ over corresponding quantization intervals.

\textit{Remarks.}
Joint density modeling of the hyperpriors, as opposed to assuming their independence, offers several advantages.
First, it enables a more accurate entropy model for estimating the true joint distribution $p(\mathbf{\tilde{y}}_1, \mathbf{\tilde{y}}_2,\mathbf{\tilde{z}}_1,\mathbf{\tilde{z}}_2)$, thereby improving rate-distortion performance.
Specifically, the joint entropy model is defined as $p(\mathbf{\tilde{y}}_1, \mathbf{\tilde{y}}_2,\mathbf{\tilde{z}}_1,\mathbf{\tilde{z}}_2)=p(\mathbf{\tilde{z}}_1,\mathbf{\tilde{z}}_2)p(\mathbf{\tilde{y}}_1|\mathbf{\tilde{z}}_1)p(\mathbf{\tilde{y}}_2|\mathbf{\tilde{z}}_2)$, instead of assuming independent hyperpriors $p(\mathbf{\tilde{z}}_1)p(\mathbf{\tilde{z}}_2)p(\mathbf{\tilde{y}}_1|\mathbf{\tilde{z}}_1)p(\mathbf{\tilde{y}}_2|\mathbf{\tilde{z}}_2)$ as in\cite{li2024content}, which does not account for source correlations between the users.
Second, since both transmitters share this joint probability function, they can predict each other's transmission rate and adjust encoding accordingly.
Specifically, given $\mathbf{z}_1$ and the joint probability function, User 1 can estimate $\mathbf{z}_2$ using the MMSE estimator, denoted as $\mathbf{z}_2^\ast$. 
By applying $h_{s_2}(\cdot)$, User 1 can obtain an estimate of $\boldsymbol{\sigma}_2$ and $\boldsymbol{\mu}_2$, thus an estimate of the code rate of $\mathbf{y}_2$, which we denote as $-\log p(\mathbf{y}_2|\mathbf{z}_2^\ast)$. 
This estimated code rate is then incorporated in the the JSCC encoder $f_{e_1}$, facilitating the encoding of $\mathbf{y}_1$.

\section{Neural Network Architecture Design}

This section presents the detailed NN architectures.

\subsection{Nonlinear Transform Modules}

Following \cite{dai2022nonlinear}, we adopt the Swin Transformer architecture \cite{liang2021swinir} as the backbone of our D-NTSCC.
Fig.~\ref{ntc modules}(a) shows the NN architectures of nonlinear transforms $g_{a_i}$ and $g_s$.
In $g_{a_i}$, each image passes through four stages of transformer blocks, gradually downsampled into a latent representation $\mathbf{y}_i\in\mathbb{R}^{\frac{H}{16}\times\frac{W}{16}\times C_4}$.
Conversely, the function $g_s$ upsamples the concatenated latent representations $\mathbf{\hat{y}}_1\oplus\mathbf{SI}_{2\rightarrow 1}\in\mathcal{R}^{\frac{H}{16}\times\frac{W}{16}\times 2C_4}$ and $\mathbf{\hat{y}}_2\oplus\mathbf{SI}_{1\rightarrow 2}\in\mathcal{R}^{\frac{H}{16}\times\frac{W}{16}\times 2C_4}$ into $\mathbf{\hat{x}}_1$ and $\mathbf{\hat{x}}_2$, respectively.
The NN architectures for the analysis and synthesis transforms of the hyperprior $\mathbf{z}_i$ are summarized in Fig.~\ref{ntc modules}(b).
These architectures consist of convolutional layers combined with ReLU activation functions.

\begin{figure*}[t]
    \centering
    \begin{subfigure}[b]{0.32\textwidth}
        \centering
        \includegraphics[width=\textwidth]{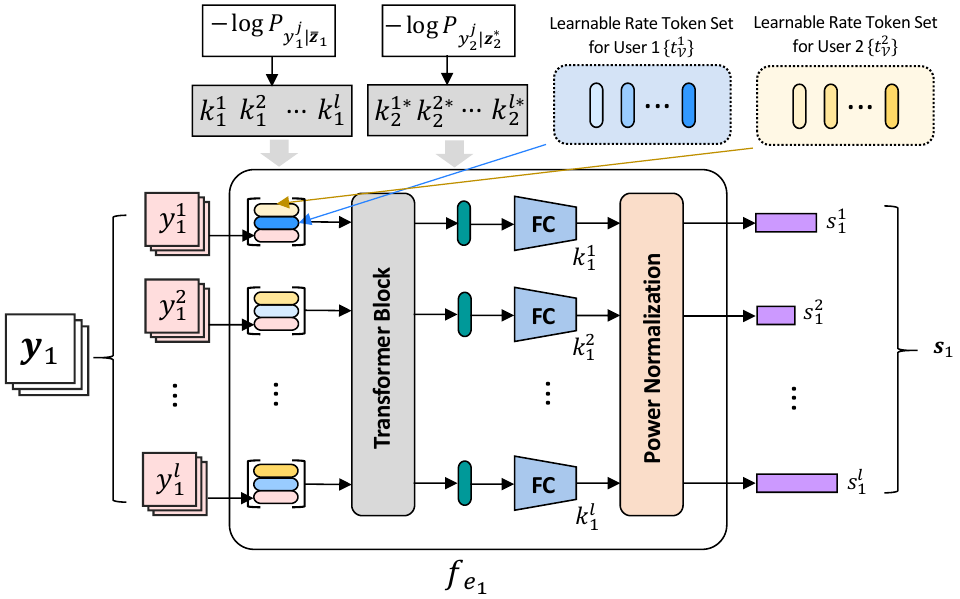}
        \caption{JSCC encoder $f_{e_1}$.}
        \vspace{-0.2cm}
    \end{subfigure}
    \begin{subfigure}[b]{0.32\textwidth}
        \centering
        \includegraphics[width=0.9\textwidth]{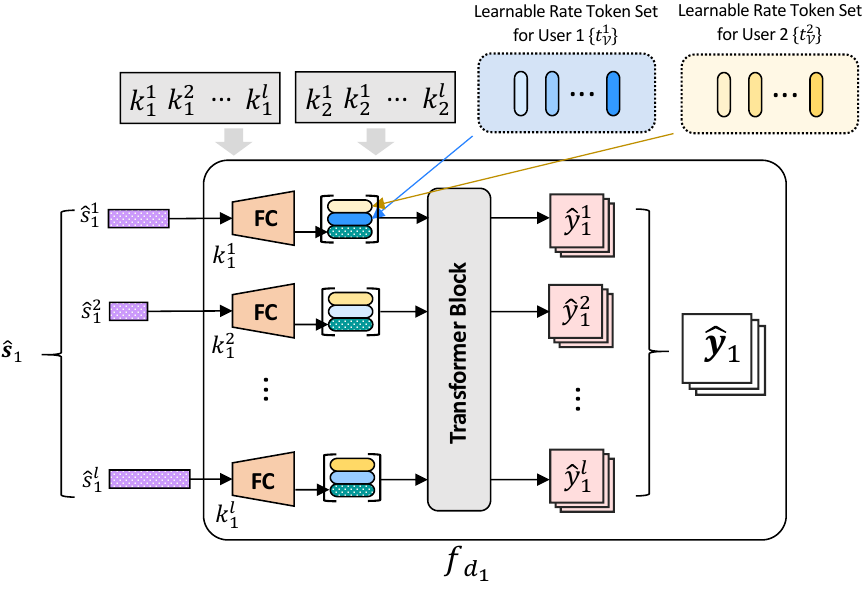}
        \caption{JSCC decoder $f_{d_1}$.}
        \vspace{-0.2cm}
    \end{subfigure}
    \begin{subfigure}[b]{0.32\textwidth}
        \centering
        \includegraphics[width=0.7\textwidth]{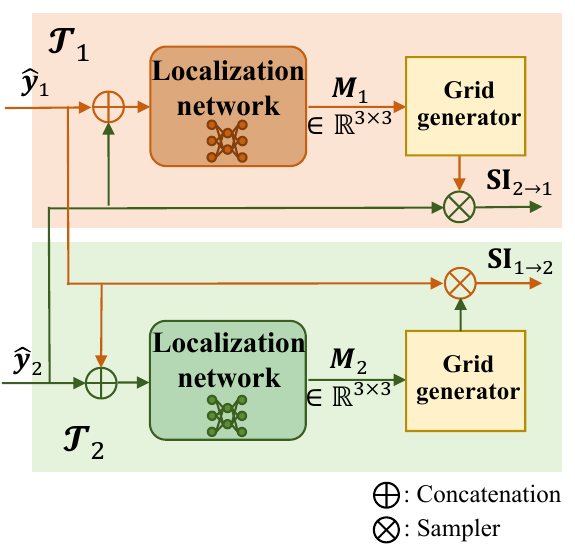}
        \caption{Transformation module.}
        \vspace{-0.2cm}
    \end{subfigure}
    \caption{NN architectures of different modules.}
    \label{jscc modules}
    \vspace{-0.6cm}
\end{figure*}

\subsection{JSCC modules}

The JSCC encoder $f_{e_i}$ at the transmitter further compresses $\mathbf{y}_i$ according to its code rate, generating the channel input symbols $\mathbf{s}_i$, as illustrated in Fig. \ref{jscc modules}(a).
Specifically, for User 1, the latent representation $\mathbf{y}_1$ is first reshaped to the size of $l\times C_4$, where $l=\frac{H}{16}\cdot\frac{W}{16}$.
Thus, $\mathbf{y}_1$ can be written as $\mathbf{y}_1=(y_1^1, y_1^2, ..., y_1^l)$, where each element has a length of $C_4$.
The entropy sum of $y_1^j$, for $j=1,...,l$, denoted as $-\log P_{y_1^j|\mathbf{\bar{z}}_1}$, is used to determine the bandwidth cost of $s_1^j$, denoted as $k_1^j$.
The value $k_1^j$ belongs to a set of integers $\mathcal{V}=\{v_1, v_2, ..., v_q\}$, and is determined by
\begin{equation}
\setlength\abovedisplayskip{4pt}
\setlength\belowdisplayskip{5pt}
    k_1^j = \underset{v\in\mathcal{V}}{\mathrm{argmin}}\left\|-\eta \log P_{y_1^j|\mathbf{\bar{z}}_1} - v \right\|.
\end{equation}
The JSCC encoder in the proposed D-NTSCC extends the one in \cite{dai2022nonlinear} by additionally incorporating the statistics of the other user.
The estimated entropy sum of $y_2^j$, denoted as $-\log P_{y_2^j|\mathbf{z}_2^\ast}$, is utilized to compute the estimated bandwidth cost of $s_2^j$, denoted as $k_2^{j\ast}$.
The encoder adapts its output based not only on its own rate information but also on the rate statistics of the other user. 
To indicate the rate information, two learnable rate token sets, $\{t_\mathcal{V}^1\}$ and $\{t_\mathcal{V}^2\}$, are introduced for User 1 and User 2, respectively.
The corresponding tokens are selected and concatenated with $y_1^j$, which is then passed into a Transformer block, leveraging its self-attention mechanism to effectively integrate the rate information.
Then, a set of variable-length fully connected (FC) layers encodes $y_1^j$ into the desired dimensions, resulting in $\mathbf{s}_1$ with variable-length elements.

At the receiver, the JSCC decoder processes $\mathbf{\hat{s}}_i$ into the recovered latent representation $\mathbf{\hat{y}}_i\in\mathbb{R}^{l\times C_4}$.
This process serves as the reverse operation of the JSCC encoder, as illustrated in Fig.~\ref{jscc modules}(b).

\subsection{Transformation Module for Alignment}

The proposed transformation module aims to align the two received latent representations prior to their concatenation, thereby enabling $g_s$ to better exploit their correlation.
We seek to achieve this goal by using spatial transformers (STs)\cite{stn}, which spatially transform input latent representations in a learnable and differentiable manner.

Fig.~\ref{jscc modules}(c) illustrates the architecture of the transformation module.
Specifically, $\mathcal{T}_1$ and $\mathcal{T}_2$ take concatenated inputs of $\mathbf{\hat {y}}_1$ and $\mathbf{\hat {y}}_2$.
These concatenated inputs are first passed through a localization network, consisting of convolutional layers and fully connected (FC) layers, to produce a pointwise projective transformation matrix $\mathbf{M}_i \in \mathbb{R}^{3\times 3}$ ($i \in \{1, 2\}$).
Here, \textit{pointwise} indicates that the same projective transformation is applied to each element (each ``pixel'') of the received latent representation.
The matrices $\mathbf{M}_1$ and $\mathbf{M}_2$ are then used by the grid generator to create sampling grids for $\mathbf{\hat{y}}_2$ and $\mathbf{\hat {y}}_1$, respectively, specifying sampling locations to produce $\mathbf{SI}_{2\to 1}$ and $\mathbf{SI}_{1\to 2}$.
A sampler applies the transformation by sampling the corresponding $\mathbf{\hat {y}}_i$ at the specified grid points, using methods such as bilinear interpolation\cite{stn}.
This whole process is differentiable, allowing for the parameter updates of the localization network through standard stochastic gradient descent (SGD) algorithms.

\begin{figure*}
     \centering
     \begin{subfigure}[b]{0.4\textwidth}
         \centering
         \includegraphics[scale=0.21]{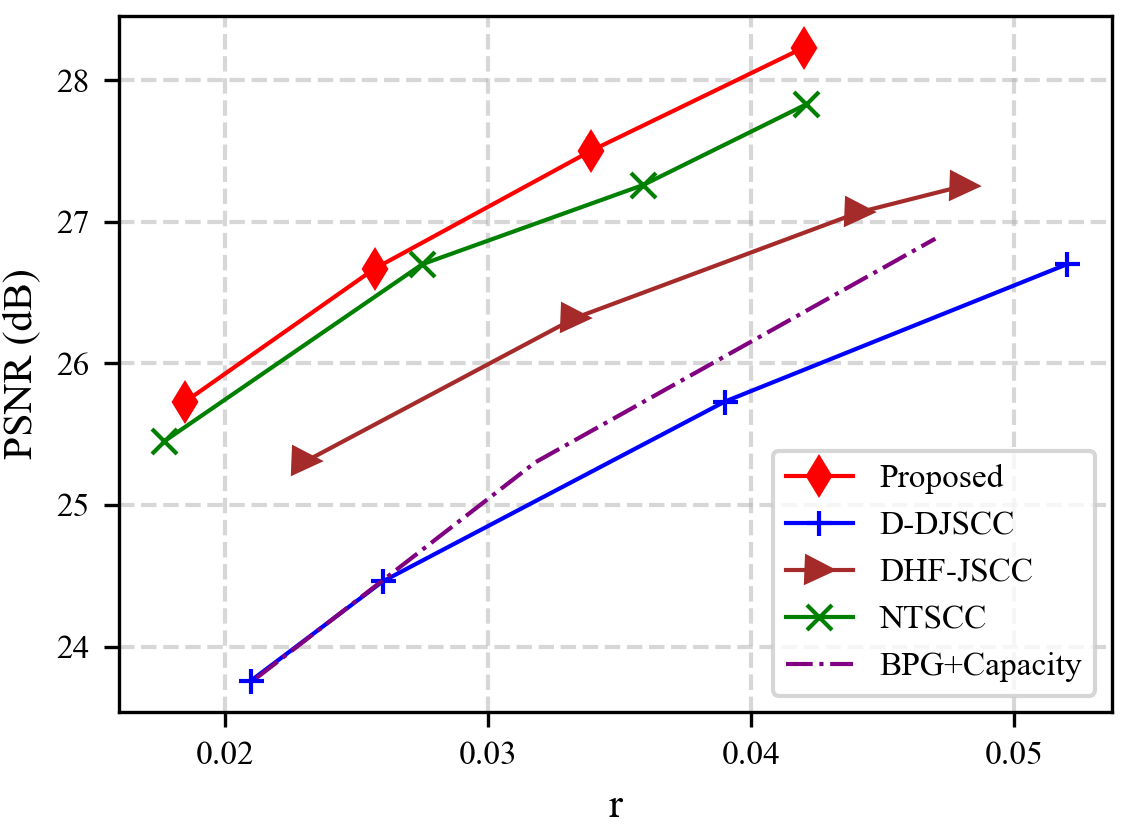}
         \caption{PSNR performances.}
         \vspace{-0.2cm}
     \end{subfigure}
     \hfill
     \begin{subfigure}[b]{0.4\textwidth}
         \centering
         \includegraphics[scale=0.21]{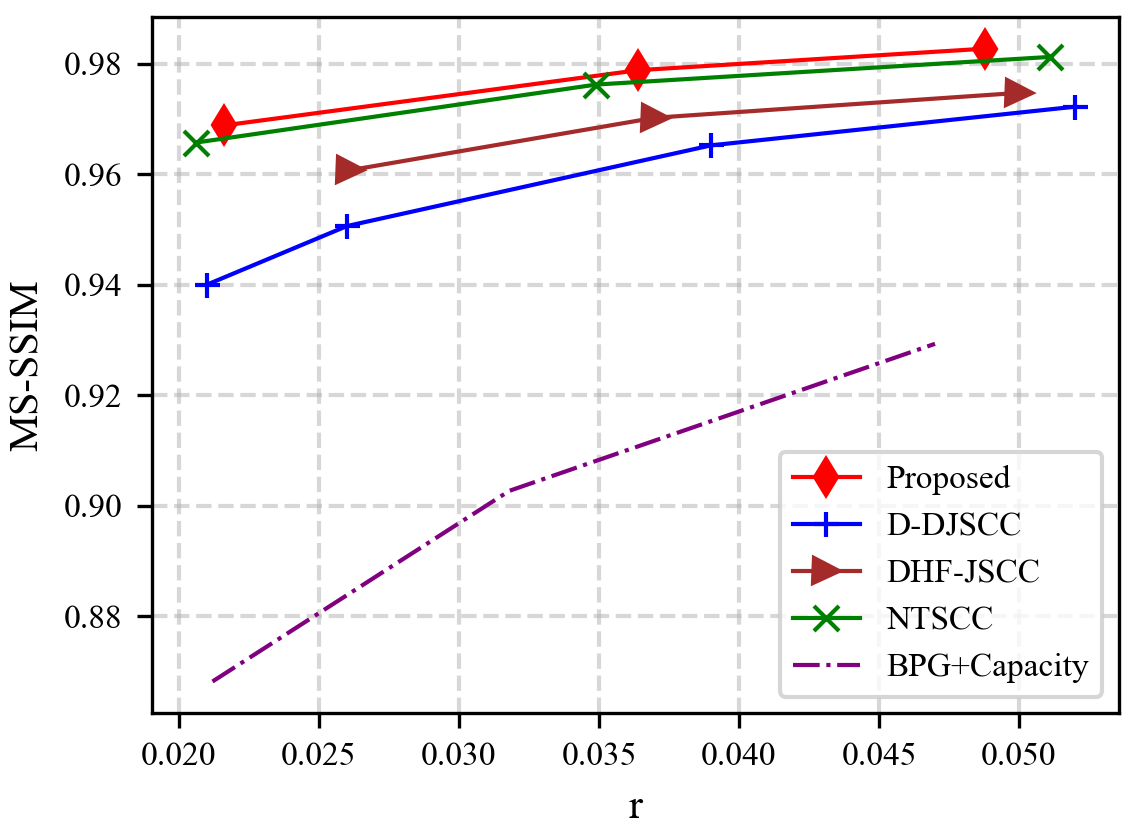}
         \caption{MS-SSIM performances.}
         \vspace{-0.2cm}
     \end{subfigure}
     \hfill
        \caption{Performances at varying transmission rates when channel SNR = 5 dB.}
        \label{snr}
        \vspace{-0.6cm}
\end{figure*}


\section{Experimental Results}


\subsection{Experiment settings}

\subsubsection{Dataset}
Our experiments are conducted on the \textit{KITTI Stereo} dataset, which consists of 1578 training stereo image pairs, and 790 testing stereo image pairs.
A stereo image pair means a pair of images each taken at the same time from a different camera.
All images are first center-cropped to a size of $370\times 740$, then downsampled to $128\times256$.

\subsubsection{Training details}
We employ the Adam optimizer for the training, and use a batch size of 2.
For the learning rate (LR), we use a cosine annealing schedule with an initial LR of $1\times10^{-4}$, which in the duration of 200 epochs gradually decreases to $1\times10^{-6}$ according to $lr(t)=10^{-6}+\frac{1}{2}(10^{-4}-10^{-6})(1+\cos\frac{t}{200}\pi)$, where $t$ represents the current epoch number.
During training, the channel SNR is fixed at 5 dB for both users.

\subsubsection{Benchmarks}
We compare the performance of our proposed D-NTSCC scheme with the following benchmarks. 
\emph{(a) D-DJSCC \cite{wang2022distributed}}: 
This distributed deep JSCC scheme exploits the image correlation via a CA module.
\emph{(b) DHF-JSCC \cite{li2024content}}: 
This scheme introduces an information fusion module based on a multi-layer CA mechanism, which fuses image features at different pixel levels to exploit the correlation.
\emph{(c) NTSCC \cite{dai2022nonlinear}}:
This is the NTSCC scheme designed for point-to-point communication.
It serves as an ablation study to demonstrate the proposed scheme's ability to leverage source correlation.
Specifically, it is globally trained on all available images, regardless of camera view, and applied to both sources for local encoding and decoding.
\emph{(d) BPG+Capacity}:
This scheme is the conventional separate source and channel coding (SSCC) method which assumes a capacity-achieving channel code and does not consider source correlation.

\subsubsection{Metrics}
Two performance metrics are considered. 
One is the peak signal-to-noise ratio (PSNR), where the distortion measure in the loss function \eqref{loss} is set to be the mean-square-error (MSE). 
The other is the multi-scale structural similarity index (MS-SSIM), where the distortion measure is set to be $1-\text{MS-SSIM}$. 

\subsubsection{Transmission Rate}
The transmission rate $r$ in channel use/pixel for one user is defined as
\begin{align}
    r = \frac{1}{CHW}\left [n + \frac{1}{2\times Capacity}\mathbb{E}_{p(\mathbf{\tilde{z}}_1, \mathbf{\tilde{z}}_2)}[-\log p(\mathbf{\tilde{z}}_1, \mathbf{\tilde{z}}_2)]\right ], \nonumber
\end{align}
where ``Capacity'' refers to the channel capacity.
The first term represents the rate for transmitting the channel input symbols $\mathbf{s}_i$, and the second term corresponds to the rate for transmitting the hyperpriors using capacity-achieving channel codes.

\subsection{Performances at varying transmission rates}


Fig.~\ref{snr}(a) illustrates the PSNR performance of different schemes at varying transmission rates.
Note that since the performance of the two users is nearly identical for all schemes, we use the performance of User 1 for illustration without loss of generality.
The results show that the proposed scheme outperforms all benchmarks across all evaluated transmission rates.
Specifically, compared to the two existing distributed coding schemes, D-DJSCC and DHF-JSCC, the proposed D-NTSCC achieves up to 50$\%$ and 30$\%$ bandwidth savings, respectively.
This performance gain is attributed to the joint entropy model and the ``upgrade'' of the NN architecture from CNN to Swin Transformer.
Additionally, the proposed scheme achieves around 10$\%$ bandwidth savings compared to the point-to-point NTSCC scheme, demonstrating its effectiveness in exploiting the source correlation.


Similar trends to those observed in the PSNR performance can be seen in the MS-SSIM curves shown in Fig.~\ref{snr}(b).
The D-NTSCC scheme achieves the highest MS-SSIM performance across all evaluated transmission rates, resulting in bandwidth savings of 16$\%$, 50$\%$, and 38$\%$ compared to NTSCC, D-DJSCC, and DHF-JSCC, respectively.

\subsection{Computational Complexity}

\begin{table}[t]
\renewcommand{\arraystretch}{1.3}
\centering
\caption{Computational complexity of different methods.}
\begin{tabular}{ccc}
\hline
Method                & FLOPs & Params \\ \hline
D-NTSCC               & 69.07 G     & 70.19 M      \\ 
NTSCC                 & 28.35 G     & 44.04 M      \\ 
D-DJSCC               & 60.79 G    & 13.35 M      \\ 
DHF-JSCC & 163.68 G   & 25.37 M      \\ \hline
\end{tabular}
\vspace{-0.6cm}
\label{computation}
\end{table}

\begin{figure*}
    \begin{small}
    \begin{align}
        &\ \ \ \ \ \mathbb{E}_{p_{\mathbf{x}_1,\mathbf{x}_2}}D_{KL}[q(\mathbf{\hat{s}}_1,\mathbf{\hat{s}}_2,\mathbf{\tilde{z}}_1,\mathbf{\tilde{z}}_2|\mathbf{x}_1,\mathbf{x}_2)||p(\mathbf{\hat{s}}_1,\mathbf{\hat{s}}_2,\mathbf{\tilde{z}}_1,\mathbf{\tilde{z}}_2|\mathbf{x}_1,\mathbf{x}_2)]\nonumber\\
        &\overset{(a)}{=}\!\mathbb{E}_{p_{\mathbf{x}_1,\mathbf{x}_2}}\mathbb{E}_{q_{\mathbf{\hat{s}}_1,\mathbf{\hat{s}}_2,\mathbf{\tilde{z}}_1,\mathbf{\tilde{z}}_2|\mathbf{x}_1,\mathbf{x}_2}}\! \left\{ \log q(\mathbf{\hat{s}}_1,\mathbf{\hat{s}}_2,\mathbf{\tilde{z}}_1,\mathbf{\tilde{z}}_2|\mathbf{x}_1,\mathbf{x}_2)-\log p(\mathbf{x}_1,\mathbf{x}_2|\mathbf{\hat{s}}_1,\mathbf{\hat{s}}_2,\mathbf{\tilde{z}}_1,\mathbf{\tilde{z}}_2)-\log p(\mathbf{\hat{s}}_1,\mathbf{\hat{s}}_2,\mathbf{\tilde{z}}_1,\mathbf{\tilde{z}}_2)\right\}+\overset{\text{\small const}_1}{\boxed{\mathbb{E}_{p_{\mathbf{x}_1,\mathbf{x}_2}}
        \log p_{\mathbf{x}_1,\mathbf{x}_2}(\mathbf{x}_1,\mathbf{x}_2)}} \nonumber\\
        &\overset{(b)}{=}\!\mathbb{E}_{p_{\mathbf{x}_1,\mathbf{x}_2}}\mathbb{E}_{q_{\mathbf{\hat{s}}_1,\mathbf{\hat{s}}_2,\mathbf{\tilde{z}}_1,\mathbf{\tilde{z}}_2|\mathbf{x}_1,\mathbf{x}_2}}\! \left\{ \log q(\mathbf{\hat{s}}_1,\mathbf{\hat{s}}_2,\mathbf{\tilde{z}}_1,\mathbf{\tilde{z}}_2|\mathbf{x}_1,\mathbf{x}_2)-\log p(\mathbf{x}_1,\mathbf{x}_2|\mathbf{\hat{s}}_1,\mathbf{\hat{s}}_2,\mathbf{\tilde{z}}_1,\mathbf{\tilde{z}}_2)-\log p(\mathbf{\hat{s}}_1,\mathbf{\hat{s}}_2|\mathbf{\tilde{z}}_1,\mathbf{\tilde{z}}_2)-\log p(\mathbf{\tilde{z}}_1,\mathbf{\tilde{z}}_2) \right\}+\text{\small const}_1\nonumber\\
        &\overset{(c)}{=}\!\overset{\text{\small const}_2}{\boxed{\mathbb{E}_{p_{\mathbf{x}_1,\mathbf{x}_2}}\mathbb{E}_{q_{\mathbf{\hat{s}}_1,\mathbf{\hat{s}}_2,\mathbf{\tilde{z}}_1,\mathbf{\tilde{z}}_2|\mathbf{x}_1,\mathbf{x}_2}}\!  \log q(\mathbf{\hat{s}}_1,\mathbf{\hat{s}}_2,\mathbf{\tilde{z}}_1,\mathbf{\tilde{z}}_2|\mathbf{x}_1,\mathbf{x}_2)}}\nonumber
        \\&\ \ \ \ \ +\mathbb{E}_{p_{\mathbf{x}_1,\mathbf{x}_2}}\mathbb{E}_{q_{\mathbf{\hat{s}}_1,\mathbf{\hat{s}}_2,\mathbf{\tilde{z}}_1,\mathbf{\tilde{z}}_2|\mathbf{x}_1,\mathbf{x}_2}}\left\{-\log p(\mathbf{x}_1,\mathbf{x}_2|\mathbf{\hat{s}}_1,\mathbf{\hat{s}}_2,\mathbf{\tilde{z}}_1,\mathbf{\tilde{z}}_2)-\log p(\mathbf{\hat{s}}_1|\mathbf{\tilde{z}}_1)-\log p(\mathbf{\hat{s}}_2|\mathbf{\tilde{z}}_2)-\log p(\mathbf{\tilde{z}}_1,\mathbf{\tilde{z}}_2) \right\}+\text{\small const}_1\nonumber\\
        &\overset{(d)}{=}\!\mathbb{E}_{p_{\mathbf{x}_1,\mathbf{x}_2}}\mathbb{E}_{q_{\mathbf{\hat{s}}_1,\mathbf{\hat{s}}_2,\mathbf{\tilde{z}}_1,\mathbf{\tilde{z}}_2|\mathbf{x}_1,\mathbf{x}_2}}\! \left\{ -\log p(\mathbf{x}_1,\mathbf{x}_2|\mathbf{\hat{s}}_1,\mathbf{\hat{s}}_2,\mathbf{\tilde{z}}_1,\mathbf{\tilde{z}}_2)-\log p(\mathbf{\hat{s}}_1|\mathbf{\tilde{z}}_1)-\log p(\mathbf{\hat{s}}_2|\mathbf{\tilde{z}}_2)-\log p(\mathbf{\tilde{z}}_1,\mathbf{\tilde{z}}_2) \right\}+\text{\small const}
        \label{expansion}
    \end{align}
    \begin{equation}
    q(\mathbf{\hat{s}}_1,\mathbf{\hat{s}}_2,\mathbf{\tilde{z}}_1,\mathbf{\tilde{z}}_2|\mathbf{x}_1,\mathbf{x}_2)=\prod_k\mathcal{N}(\hat{s}_1^k|s_1^k,\epsilon_1^2)\prod_j\mathcal{N}(\hat{s}_2^j|s_2^j,\epsilon_2^2)\prod_m\mathcal{U}(z_1^m|z_1^m-\frac{1}{2},z_1^m+\frac{1}{2})\prod_n\mathcal{U}(\tilde{z}_2^n|z_2^n-\frac{1}{2},z_2^n+\frac{1}{2})
    \label{q}
    \end{equation}
    \hrulefill
    \end{small}
    \vspace{-0.6cm}
\end{figure*}

Table \ref{computation} compares the computational complexity of different methods.
As can be seen, while D-NTSCC outperforms NTSCC, it comes at the cost of increased computational complexity, requiring more floating point operations (FLOPs) and memory.
Other CNN-based methods consume less memory than D-NTSCC. 
In terms of FLOPs, the proposed method increases FLOPs by 13.6$\%$ compared to D-DJSCC, but only requires 42.2$\%$ of the FLOPs of DHF-JSCC.
Overall, D-NTSCC offers a reasonable trade-off between performance and computational complexity, making it a potential candidate for practical applications.

\section{Conclusion}
This work introduces D-NTSCC, a deep learning-based distributed source-channel coding approach for correlated image transmission. 
The scheme separately encodes two correlated image sources at the transmitters and jointly decodes them at the receiver.
By proposing a learned joint entropy model, this scheme can better capture the dependencies between the sources, thereby improving rate-distortion performance.
The corresponding loss function is further derived to jointly optimize the encoding, decoding, and the joint entropy model, ensuring accurate joint distribution approximation.
Experiments demonstrate D-NTSCC's advantages over existing distributed coding schemes, highlighting its potential for distributed semantic communications.

\begin{appendices}
\section{Proof of Proposition \ref{prop}}

To prove proposition 1, we begin by expanding the KL divergence as shown in \eqref{expansion}.
Equation (a) follows directly from the definition of the KL divergence.
The last term in (a) is a constant, as the source probability remains fixed.
Equation (b) applies the multiplication rule of probability.
Equation (c) holds due to the Markov chain $\mathbf{\hat{s}}_1-\mathbf{\tilde{z}}_1-\mathbf{\tilde{z}}_2-\mathbf{\hat{s}}_2$, which leads to the factorization $p(\mathbf{\hat{s}}_1,\mathbf{\hat{s}}_2|\mathbf{\tilde{z}}_1,\mathbf{\tilde{z}}_2)=p(\mathbf{\hat{s}}_1|\mathbf{\tilde{z}}_1)p(\mathbf{\hat{s}}_2|\mathbf{\tilde{z}}_2)$.
Moreover, the first term in (c) is a constant, since given $\mathbf{x}_1$ and $\mathbf{x}_2$, $q(\mathbf{\hat{s}}_1,\mathbf{\hat{s}}_2,\mathbf{\tilde{z}}_1,\mathbf{\tilde{z}}_2|\mathbf{x}_1,\mathbf{x}_2)$ can be expanded as the product of known probabilities, as shown in \eqref{q}.
This leads to Equation (4), where $\text{const}=\text{const}_1+\text{const}_2$.

\end{appendices}





\bibliographystyle{IEEEtran}
\bibliography{dntscc.bib}{}

\end{document}